\documentclass[twocolumn]{aastex631}

\def\al{\alpha}
\def\be{\beta}
\def\ga{\gamma}
\def\de{\delta}

\def\th{\theta}

\def\ka{\kappa}
\def\la{\lambda}

\def\si{\sigma}

\def\ph{\phi}

\def\mn{{\mu\nu}}
\def\abgd{{\al\be\ga\de}}

\def\lsim{\mathrel{\rlap{\lower4pt\hbox{\hskip1pt$\sim$}}
    \raise1pt\hbox{$<$}}}
\def\gsim{\mathrel{\rlap{\lower4pt\hbox{\hskip1pt$\sim$}}
    \raise1pt\hbox{$>$}}}
\def\sqr#1#2{{\vcenter{\vbox{\hrule height.#2pt
         \hbox{\vrule width.#2pt height#1pt \kern#1pt
         \vrule width.#2pt}
         \hrule height.#2pt}}}}

\def\lrpartial{\raise 1pt\hbox{$\stackrel\leftrightarrow\partial$}}

\def\part2{\partial_\alpha \partial^\alpha}

\def\pt#1{\phantom{#1}}

\def\tt{$t^{\ka\la\mu\nu}$}

\def\xx'{|\vec x -\vec x'|}

\def\b2{b^\al b_\al}

\newcommand{\beq}{\begin{equation}}
\newcommand{\eeq}{\end{equation}}
\newcommand{\bea}{\begin{eqnarray}}
\newcommand{\eea}{\end{eqnarray}}
\newcommand{\bit}{\begin{itemize}}
\newcommand{\eit}{\end{itemize}}
\newcommand{\rf}[1]{(\ref{#1})}
\newcommand\bw{\begin{widetext}}
\newcommand\ew{\end{widetext}}
\submitjournal{ApJ}

\shorttitle{Bumblebee black holes}
\shortauthors{Xu et al.}

\begin{document}

\title{Bumblebee black holes in light of Event Horizon Telescope observations}

\correspondingauthor{Rui Xu and Lijing Shao}
\email{xuru@pku.edu.cn, lshao@pku.edu.cn}

\author[0000-0002-7157-3805]{Rui Xu}
\affiliation{Kavli Institute for Astronomy and Astrophysics, Peking University, Beijing 100871, China}

\author[0000-0001-5021-235X]{Dicong Liang}
\affiliation{Kavli Institute for Astronomy and Astrophysics, Peking University, Beijing 100871, China}

\author[0000-0002-1334-8853]{Lijing Shao}
\affiliation{Kavli Institute for Astronomy and Astrophysics, Peking University, Beijing 100871, China}
\affiliation{National Astronomical Observatories, Chinese Academy of Sciences, Beijing 100012, China}



\begin{abstract}

We report the existence of novel static spherical black-hole solutions in a
vector-tensor gravitational theory called the bumblebee gravity model which
extends the Einstein-Maxwell theory by allowing the vector to nonminimally
couple to the Ricci curvature tensor.  A test of the solutions in the
strong-field regime is performed for the first time using the recent
observations of the supermassive black-hole shadows in the galaxy M87 and the
Milky Way from the Event Horizon Telescope Collaboration. The parameter space is
found largely unexcluded and more experiments are needed to fully bound the
theory.

\end{abstract}



\section{Introduction} \label{sec:intro}
The bumblebee gravity model is a vector-tensor gravitational theory that serves
as an alternative to Einstein's general relativity (GR). The vector field in the
theory, called the bumblebee field, enters the action the same way as the
electromagnetic (EM) four-potential, only that the bumblebee field $B_\mu$
couples to the Ricci tensor $R_\mn$ quadratically and might acquire a nontrivial
background configuration due to a cosmological potential term $V$. The action of
the bumblebee model reads \citep{Kostelecky:2003fs}
\bea
S &=& \int d^4x \sqrt{-g} \left( \frac{1}{2\ka} R + \frac{\xi}{2\ka} B^\mu B^\nu
R_\mn - \frac{1}{4} B^\mn B_\mn - V \right) 
\nonumber \\
& & + S_{\rm m} ,
\label{actionB}
\eea  
where $g$ is the determinant of the metric, $\ka=8\pi G$ with $G$ being the
gravitational constant and set to unity in the remaining of the work,
$R$ is the Ricci scalar, and $S_{\rm m}$ is the action of conventional matter.
The EM-like field strength is defined as $B_\mn := D_\mu B_\nu - D_\nu B_\mu$
with $D_\mu$ being the covariant derivative. The coupling term $B^\mu B^\nu
R_\mn$ controlled by the constant $\xi$ indicates that when $\xi \ne 0$, the
bumblebee vector field nonminimally couples to the metric tensor, generating
more sophisticated solutions than those in the Einstein-Maxwell theory.

The bumblebee gravity model has been well investigated in the weak-field regime
\citep{Kostelecky:2003fs, Bluhm:2004ep, Bluhm:2005uj, Bailey:2006fd,
Bluhm:2007bd, Bluhm:2008yt, Liang:2022hxd}, and widely used as an example for
theories where spontaneous violation of Lorentz symmetry occurs as the bumblebee
field acquires a certain background configuration that minimizes the
cosmological potential $V$. Because the form of $V$ is unknown, the practice in
studying violation of Lorentz symmetry usually assumes a constant background
configuration of the bumblebee field in asymptotically flat regions. Then the
constant background bumblebee field can be related to the coefficients for
Lorentz-symmetry violation in the popular framework of the Standard-Model
Extension (SME) \citep{Colladay:1996iz, Colladay:1998fq, Kostelecky:2003fs},
which collects all possible Lorentz-violating terms at the level of effective
field theory \citep{Kostelecky:2009zp, Kostelecky:2011gq, Kostelecky:2013rta,
Kostelecky:2018yfa} and predicts effects to confront with experiments
\citep{Kostelecky:2008ts, Tasson:2016xib}, illustrating and complementing
assumptions and conclusions in the general SME framework. 

While the study of the bumblebee model in the weak-field regime leads to
stringent constraints on the bumblebee field in asymptotically flat regions, it
becomes more and more essential to obtain strong-field solutions so that the
model as well as Lorentz symmetry can be tested in the strong-field regime using
the state-of-the-art astrophysical observations from gravitational-wave (GW)
observatories \citep{LIGOScientific:2018mvr, LIGOScientific:2020ibl,
LIGOScientific:2021djp} and the black-hole shadows
\citep{EventHorizonTelescope:2019dse, EventHorizonTelescope:2022xnr}. As a first
step to this destination, we report newly found numerical solutions of black
holes in the bumblebee model and consequently utilize the shadows of the two
supermassive black holes constructed by the Event Horizon Telescope (EHT)
Collaboration to constrain the bumblebee field in the vicinity of the black
holes. The numerical solutions reported here extend the Reissner-Nordstr\"om
solution as the bumblebee model can be regarded as an extension of the
Einstein-Maxwell theory. A charge similar to the electric charge can be defined
for the bumblebee field accompanying the black holes. Our results put
simultaneous bounds on the coupling constant $\xi$ and the bumblebee charge of
the black holes for the first time.

\section{Static spherical black holes in the bumblebee model.}
\label{sec:ii}

The setup of equations and the numerical method have been described in detail in \citet{Xu:2022frb}. We briefly summarize the results here.  

The field equations can be derived from the action in Eq.~\rf{actionB} by
variations of the metric and the bumblebee field. Since we are looking for
vacuum solutions that already minimize the cosmological potential $V$, it then
merely plays the role of a cosmological constant which we will drop in current
study of black-hole solutions. Denoting the background bumblebee field as
$b_\mu$, the vacuum field equations turn out to be
\bea
&& G_\mn =  \ka \left( T_{b}\right)_\mn , 
\nonumber \\
&& D^\mu b_{\mn} + \frac{\xi}{\ka} b^\mu R_\mn  = 0,
\label{fieldeqs2}
\eea
where $G_\mn$ is the Einstein tensor and $b_\mn = D_\mu b_\nu - D_\nu b_\mu$.
The energy-momentum tensor of $b_\mu$ is 
\bea
\left( T_{b}\right)_\mn &=& \frac{\xi}{2\ka} \Big[ g_\mn b^\al b^\be R_{\al\be} - 2 b_\mu b_\la R_\nu^{\pt\nu \la} - 2 b_\nu b_\la R_\mu^{\pt\mu \la} 
\nonumber \\
&& - \Box_g ( b_\mu b_\nu ) - g_{\mn} D_\al D_\be ( b^\al b^\be ) + D_\ka D_\mu \left( b^\ka b_\nu \right) 
\nonumber \\
&& + D_\ka D_\nu ( b_\mu b^\ka )   \Big] + b_{\mu\la} b_\nu^{\pt\nu \la} - \frac{1}{4} g_\mn  b^{\al\be} b_{\al\be}   ,
\eea
with $\Box_g=g^{\al\be}D_\al D_\be$ being the d'Alembertian in the curved
spacetime.

To find static spherical black holes, the metric ansatz 
\bea
ds^2 = -e^{2\nu}dt^2 + e^{2\mu} dr^2 + r^2 \left( d\th^2 + \sin^2\th \, d\ph^2 \right) 
\label{ssmetric}
\eea
can be used. For the background bumblebee field, it can only have the temporal
component $b_t$ and the radial component $b_r$ to respect the spherical
symmetry. The unknowns $\nu, \, \mu, \, b_t$ and $b_r$ are functions of the
radial coordinate $r$ and to be solved from the field equations. Similar to the
Maxwell theory, the radial component of the bumblebee field is nondynamic due to
the specific form of the kinetic term in Eq.~\rf{actionB}. This is reflected in
the fact that $b_r$ and its derivatives can be completely eliminated in the
field equations without raising the orders of derivatives of the other
variables. However, unlike the Einstein-Maxwell theory where the energy-momentum
tensor of the EM field is gauge invariant, the energy-momentum tensor of the
bumblebee field is not. So the nondynamic radial component $b_r$, if
nonvanishing, contributes to $\left( T_{b}\right)_\mn$, becoming part of the
source for spacetime curvature.

It turns out that there are two families of vacuum solutions corresponding to
vanishing and nonvanishing $b_r$ respectively. The first family of solutions,
characterized by $b_r=0$, naturally extends the Reissner-Nordstr\"om solution
and is what we will test against the EHT observations. The second family of
solutions with nonvanishing $b_r$ is characterized by $R_{rr}=0$, and only
exists for $\xi \ne 0$, as it can be seen from the radial component of the
vector field equation in Eq.~\rf{fieldeqs2}. Let us mention that the analytic solution found in \citet{Casana:2017jkc} belongs to the second family. An extension for the result of Casana et al. is obtained in \citet{Xu:2023zjw}, and the entire second family of solutions, including numerical ones, is studied in \citet{Xu:2022frb} in detail. The fact that the bumblebee theory possesses two distinct families of spherical black-hole solutions has not been discussed in the literature to our best knowledge.

It is interesting to point out that the second family of solutions can be
regarded as a generalization of the Schwarzschild metric due to the fact that it
remarkably includes an analytic solution whose metric is exactly the
Schwarzschild metric
\bea
&& \nu = -\mu= \frac{1}{2} \ln{\left(1-\frac{2M}{r} \right)} , 
\label{schbh}
\eea
while $b_t$ and $b_r$ are given by
\bea
&& b_t = \la_{0} + \frac{\la_{1}}{r} , 
\nonumber \\
&& b_r = \pm \bigg[ \frac{\la_1^2 (2r-M) + 6\la_0\la_1 Mr + 6 \la_0^2 M^2 r}{3M (r-2M)^2} 
\nonumber \\
&& \hskip 1.1cm  - \frac{\ka \la_1^2}{3\xi M (r-2M)} \bigg]^{\frac{1}{2}},
\label{asol1}
\eea
with $\la_0$ and $\la_1$ being constant. Non-Schwarzschild metric exists in the
second family of solutions, but it happens that the leading-order effect on the
advance of perihelion for an orbit in the spacetime of these solutions deviates
from the GR result regardless of the bumblebee field. So the Solar-system
observations of the planets' orbits have severely restricted the difference
between the metric of these solutions and the Schwarzschild metric
\citep{Casana:2017jkc}. In other words, as long as the Solar-system constraints
are satisfied, the solutions in the second family converge to the analytic
solution given by Eqs.~\rf{schbh} and \rf{asol1}, so the shadow of a black hole
in this family is very close to that of a Schwarzschild black hole no matter how
large $\la_0$ and $\la_1$ are.  This is why we are not to test the second family
of solutions against the EHT observations.

Focusing on the first family of solutions where $b_r=0$, ordinary differential
equations (ODEs) for $\nu, \, \mu$ and $b_t$ from Eq.~\rf{fieldeqs2} can be
numerically integrated either from a large radius inward or from a given value
of the radius for the event horizon outward. Each black-hole solution in this
family has two independent parameters, maybe chosen as the asymptotic
quantities: the mass $M$ and the bumblebee charge $Q$ of the black hole, or as
the quantities on the event horizon: the radius of the event horizon $r_h$ and
the limit of $(r-r_h)\, g_{rr}$ when $r\rightarrow r_h$. Practically we find
that the error of the numerical solutions is easier to control when integrating
from the event horizon outward with $r_h$ and the limit of $(r-r_h)\, g_{rr}$ at
$r\rightarrow r_h$ specified than integrating from a large radius inward with the
mass and the charge specified. But for the presentation of the results in this work, let us speak of the mass $M$ and the charge $Q$ as the two
indepent parameters of the black hole as they are quantities more intuitive than
the limit of $(r-r_h)\, g_{rr}$ at $r\rightarrow r_h$.

Technically, the mass $M$ and the bumblebee charge $Q$ of the black hole are
defined to be proportional to the coefficients of the $1/r$ terms in the
asymptotic expressions for the variables $\mu$ and $b_t$ respectively. Denoting
\bea
&& \mu = \frac{\mu_1}{r} + \frac{\mu_2}{r^2} + \frac{\mu_3}{r^3} + \cdots \, ,
\nonumber \\
&& b_t = \la_0 + \frac{\la_1}{r} + \frac{\la_2}{r^2} + \frac{\la_3}{r^3} + \cdots \, ,
\label{asyexp}
\eea  
then $M:=\mu_1$ and $Q:= \sqrt{\ka/2} \, \la_1$. The other coefficients $\mu_2,
\, \mu_3, \, \cdots$ and $\la_2, \, \la_3, \, \cdots$ in Eq.~\rf{asyexp} as well
as expansion coefficients for $\nu$ are related to $\mu_1$ and $\la_1$ (or
equivalently, $M$ and $Q$) by certain recurrence relations that can be derived
from the ODEs (see Eq.~(B1) in \citet{Xu:2022frb}). Note that there is no
recurrence relation for the coefficient $\la_0$, but it depends on $\mu_1$ and
$\la_1$ due to a nontrivial boundary condition for black-hole solutions, namely
that  $g_{tt} = -e^{2\nu}$ vanishes on the event horizon. That is to say, we
find in the numerical solutions that $g_{tt}$ is not guaranteed to vanish when
$g_{rr}$ diverges at a finite radius $r_h$ which is supposed to be the event
horizon. Solutions with nonvanishing $g_{tt}$ at the finite radius where
$g_{rr}$ diverges and the curvature scalars $R, \, R^{\mn}R_{\mn}$ and
$R^{\abgd}R_{\abgd}$ are finite are discussed in \citet{Xu:2022frb}. These
solutions are not black holes, so we do not consider them here. 

\begin{figure}
 \includegraphics[width=\linewidth]{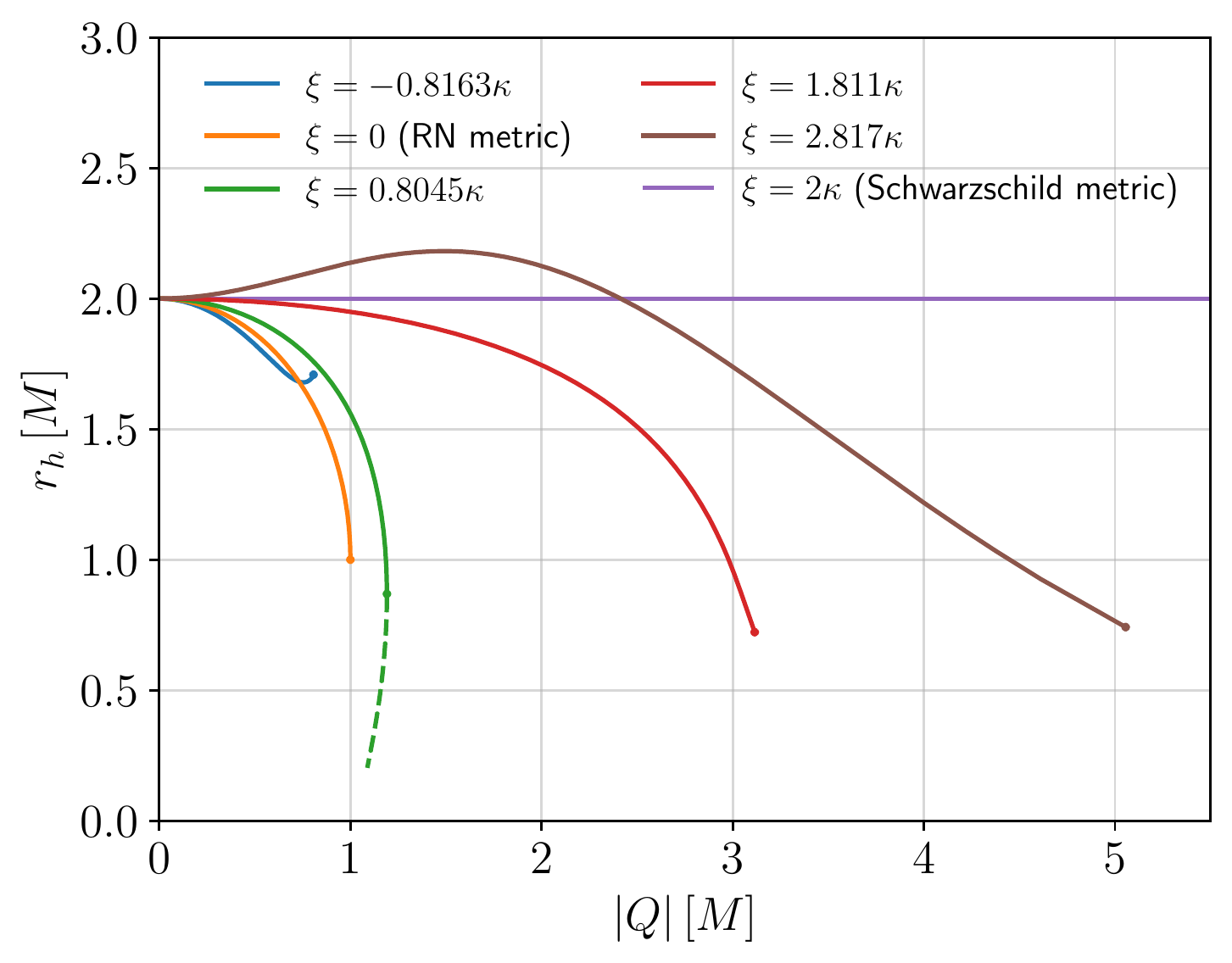}
 \caption{Radius of the event horizon $r_h$ versus bumblebee charge $Q$ of the
 black hole. Except for $\xi=2\ka$, a maximally charged black-hole solution,
 indicated by a solid dot on each line, exists for a given coupling constant
 $\xi$. Notice that for the green line with $\xi=0.8045\ka$, the bumblebee
 charge $Q$ turns back after reaching its maximum. }
\label{fig0}
\end{figure}

By specifying suitable values for $r_h$ and the limit of $(r-r_h)\, g_{rr}$ at
$r\rightarrow r_h$, we obtain black-hole solutions by integrating the field
equations outward. The mass $M$ and the charge $Q$ for each solution can then be
extracted according to Eq.~\rf{asyexp}. Using the mass $M$ as the unit, the
radius of the event horizon $r_h$ with respect to the charge $Q$ is plotted in
Fig.~\ref{fig0}, for different values of the coupling
constant $\xi$. As expected, the numerical solutions recover the
Reissner-Nordstr\"om solution when $\xi=0$. What remarkable is that when
$\xi=2\ka$, the metric of the numerical solutions becomes the Schwarzschild
metric, independent of the bumblebee charge which therefore can be arbitrarily
large. In this case, $b_t$ also has an interesting analytic expression, $b_t
\propto 1-2M/r$. We also notice that when $\xi>0$, the maximal charge-mass ratio
for the black hole is larger than unity. Finally, let us point out that when
$\xi$ is between 0 and about $1.2\ka$, two black-hole solutions exist given a
value of $Q$ near the maximal charge (see the green curve with a dashed segment
in Fig.~\ref{fig0}). We suspect the one with the smaller $r_h$ to be unstable
and drop these solutions in the following analysis though further stability
study is needed to confirm the speculation.

\begin{figure}
 \includegraphics[width=\linewidth]{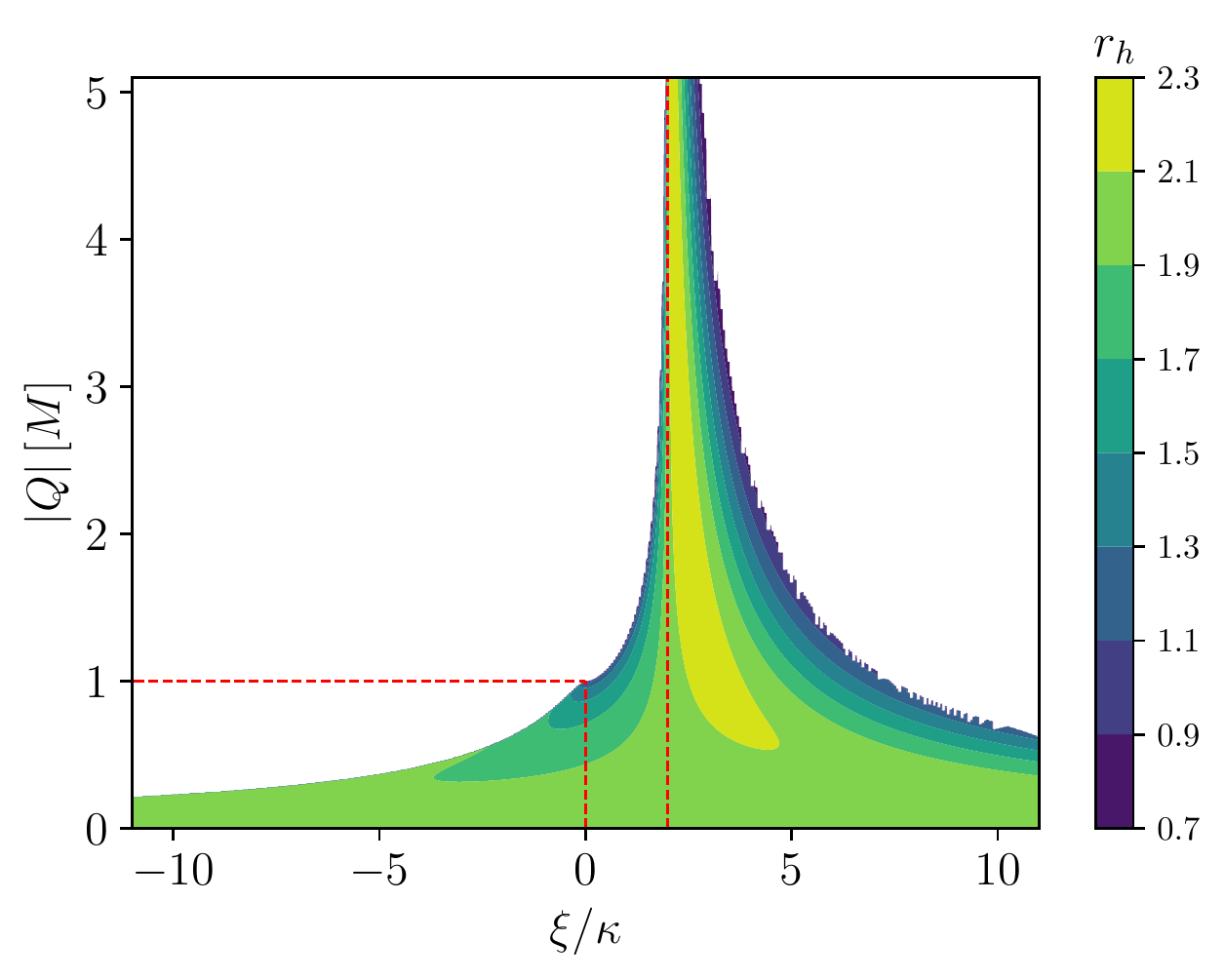}
 \caption{Contour plot for the radius of the event horizon $r_h$ on the
 $\xi$-$Q$ plane with the mass of the black hole being the unit of $r_h$ and
 $Q$. The red dashed lines indicate $\xi=0$ and $\xi=2\ka$. The uneven boundary
 on the right part of the plot is due to numerical difficulties in finding the
 maximally charged black-hole solutions when $\xi$ is large.  }
\label{fig1}
\end{figure}

To have a view of the existence domain for the black-hole solutions, we still
use the mass of the black hole as the unit, and generate in Fig.~\ref{fig1} a
contour plot representing different value levels for the radius of the event
horizon on the two-dimensional plane of the coupling constant $\xi$ and the
bumblebee charge $Q$. The boundary of the contour plot indicates the change of
the maximal charge-mass ratio with respect to the coupling constant $\xi$,
namely that black-hole solutions do not exist in the blank region. Notice that
$r_h$ does not approach zero at the boundary of the existence domain. Excluding
the suspected unstable solutions corresponding to the dashed segment in
Fig.~\ref{fig0}, the minimal radius of the event horizon for all values of $\xi$
is above $0.7M$. At $\xi=2\ka$, the metric takes the form of the Schwarzschild
solution with $r_h=2M$ regardless of the bumblebee charge $Q$ which can be
arbitrarily large. As $\xi$ goes away from the special value $\xi=2\ka$, the
maximal charge-mass ratio decreases rapidly.

\section{Constraints from the EHT observations.}
\label{sec:constraints}

The EHT Collaboration successfully constructed shadow images for the
supermassive black holes in the centers of the galaxy M87
\citep{EventHorizonTelescope:2019dse} and the Milky Way
\citep{EventHorizonTelescope:2022xnr}, providing quantitive evidences for the
scale of black-hole event horizon. When alternatives to GR predict black holes
with different sizes of event horizon, the shadow images can be used to test and
constrain them. As the shadow of a black hole is casted when the EM waves
emitted by matter near the event horizon are absorbed by the black hole, the exact size and shape
of the shadow depend not only on the size of the event horizon but also on the
distribution and the motion of matter around the black hole. The effect due to matter is complicated, but generally small when considering its correction
to the size of the shadow, given current observational uncertainties of the
masses and distances of the two supermassive black holes
\citep{EventHorizonTelescope:2019dse, EventHorizonTelescope:2022xnr}.  Therefore,
we neglect the effect due to matter in this work but acknowledge that
modelling matter surrounding the black hole is highly demanded for
improving the test as well as explaining image details.

Without concerning the distribution and the motion of matter around a black
hole, the shadow of the black hole is
determined by the trajectory of the light that barely escapes from the light
ring near the event horizon. For an observer far away from the black hole, the
radius of the shadow is the distance between the line of sight and the
approaching trajectory of the light, which is equivalently the impact parameter
for the scattering lightlike geodesic that infinitely orbits around the light
ring. In the spacetime described by the metric in Eq.~\rf{ssmetric}, lightlike
geodesics have four-velocity components
\bea
&& \frac{dt}{d\la} = e^{-2\nu} ,
\nonumber \\
&& \frac{dr}{d\la} = \pm e^{-\mu} \sqrt{ e^{-2\nu} - \frac{\si^2}{r^2} }, 
\nonumber \\
&& \frac{d\th}{d\la} = 0 ,
\nonumber \\
&& \frac{d\ph}{d\la} = \frac{\si}{r^2} ,
\eea     
where $\la$ is an affine parameter, $\si$ is an integral constant, and
spherical symmetry has been used to put the orbit in the equatorial plane. To
see that $\si$ is the impact parameter for scattering orbits, we take the
$x$-axis along the velocity direction of the orbit at infinity. Then, as $r$
goes to infinity, we have $\ph \rightarrow 0$, and the impact parameter is
\bea
\lim\limits_{r\rightarrow \infty} r \ph = \lim\limits_{r\rightarrow \infty} r^2 \times  \frac{  \big| {d\ph}/{d\la}\big| }{\big|{dr}/{d\la}\big| } = \si \lim\limits_{r\rightarrow \infty} e^{\mu+\nu} .
\eea 
The black-hole solutions to be tested here are asymptotically flat, so $\si$ is
the impact parameter.\footnote{Black-hole solutions in the second family
($b_r\ne 0$) can have nonvanishing $\mu$ at infinity though Solar-system
observations are not in favor of such solutions~\citep{Casana:2017jkc, Xu:2022frb}. }  

The value of the impact parameter $\si$ on the light ring gives the radius of
the black-hole shadow. It can be computed using the definition of the light
ring, namely      
\bea
\frac{dr}{d\la} = 0, \quad \frac{d^2r}{d\la^2} =0 .
\label{co}
\eea
The $r$-component of the four-acceleration can be obtained from the geodesic
equations. After some simplification, Eq.~\rf{co} gives
\bea
r e^{-\nu} = \si, \quad r \nu' = 1  ,
\label{co2}
\eea
where the prime denotes the derivative with respect to $r$. Our numerical
solutions are interpolated to solve the radius of the light ring, denoted as
$r_{\rm lr}$, from the second equation in Eq.~\rf{co2}. Then using $r_{\rm lr}$,
the first equation in Eq.~\rf{co2} gives the value of $\si$ for the light ring,
denoted as $\si_{\rm lr}$, which is the radius of the shadow for nonspinning
black holes. The spin of a black hole causes small distortion to the
shape of the black-hole shadow in GR, correcting deduced parameters by factors of order unity \citep{EventHorizonTelescope:2022exc, EventHorizonTelescope:2022urf}. We do not know if the same is true in bumblebee gravity as rigorous solutions of rotating black holes in the bumblebee theory have not been found to our knowledge. Though it is beyond the scope of the present work, constructing solutions of rotating black holes and putting them into test would be worthwhile as the resolution of EHT improves.\footnote{Our investigation aims at providing preliminary constraints on the bumblebee charge and the coupling constant in the theory in a simplified but sufficient way. So we focus on the size of the black-hole shadow which is described by an average diameter of the bright ring modelled by the EHT Collaboration while do not take into account the asymmetry parameter of the ring. More accurate constraints can be derived by simultaneously considering the average diameter and the asymmetry of the observationally modelled black-hole shadow.}

\begin{figure}
 \includegraphics[width=\linewidth]{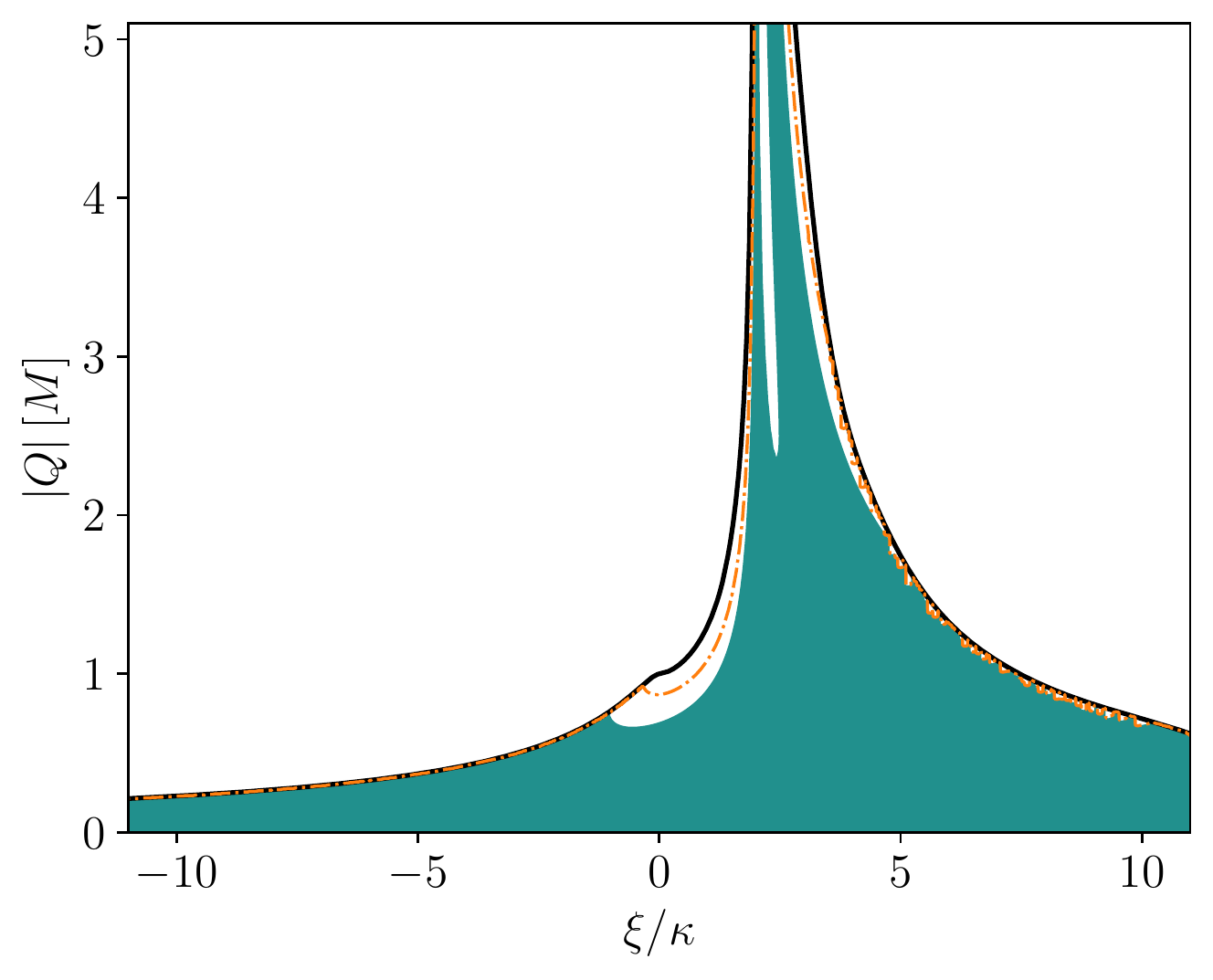}
 \caption{Constraints on the $\xi$-$Q$ plane due to the observations of the
 shadows for the supermassive black holes in M87 and the Milky Way. The shaded
 region is generated by letting numerically calculated $\si_{\rm lr}/M$ to
 satisfy the upper and lower bounds from Sgr~A* in Eq.~\rf{sag}. The dash-dotted
 orange line is by satisfying the lower bound in Eq.~\rf{m87} while the upper
 bound in Eq.~\rf{m87} is trivially satisfied by all the solutions. The solid
 black line is the smoothed boundary of the plot in Fig.~\ref{fig1}, indicating
 the existence domain of the black-hole solutions.}
\label{fig2}
\end{figure}

For observational results, we have an angular diameter $d = 42\pm 3 \, \mu {\rm
as}$ for the shadow of the black hole in M87
\citep{EventHorizonTelescope:2019dse}, and an angular diameter $d = 51.8 \pm 2.3
\, \mu{\rm as}$ for the shadow of the black hole in the Milky Way
\citep{EventHorizonTelescope:2022xnr}. As the angular diameter is related to the
radius $\si_{\rm lr}$ by
\bea
d = \frac{2\si_{\rm lr}}{D} = \frac{2\si_{\rm lr}}{M} \frac{M}{D} ,
\eea 
where $M$ and $D$ are the mass and distance of the black hole, the mass-distance
ratio is required to obtain the observational value of $\si_{\rm lr}$ in unit of
the black hole mass. For the supermassive black hole in M87, we adopt $M/D =
3.62 \pm 0.60 \, \mu {\rm as}$ from stellar dynamics \citep{Gebhardt:2011yw}, and
for the supermassive black hole in the Milky Way, we adopt the average of the
results from the Very Large Telescope Interferometer~\citep{GRAVITY:2021xju} and
Keck~\citep{Do:2019txf} while their difference is used as the uncertainty, namely
$M/D = 5.02 \pm 0.20 \, \mu {\rm as}$.\footnote{To test the theory self-consistently, the mass-distance ratios should also be deduced using the bumblebee gravity. However, since deducing the mass-distance ratios only involves using the theory in its weak-field limit where the bumblebee gravity coincides with GR, we are comfortable with using the standard results in the literature for the mass-distance ratios. We reckon that using the full theory would not introduce significant changes.} 
In conclusion, we have 
\bea
\frac{\si_{\rm lr}}{M} = 5.80 \pm 1.05 
\label{m87}
\eea
for the supermassive black hole in M87, and 
\bea
\frac{\si_{\rm lr}}{M} = 5.16 \pm 0.31
\label{sag}
\eea
for the supermassive black hole in the Milky Way. Figure \ref{fig2} shows the
region on the $\xi$-$Q$ plane for the bumblebee black holes that are consistent
with the results in Eqs.~\rf{m87} and \rf{sag}.

\section{Conclusions.}
\label{sec:disc}

In this work, we have presented black-hole solutions in the bumblebee gravity
model described by the action in Eq.~\rf{actionB}. The coupling between the
bumblebee vector field and the Ricci tensor enriches the content of the theory,
leading to two families of black-hole solutions. One extends the
Reissner-Nordstr\"om solution in the Einstein-Maxwell theory, and the other
appears to generalize the Schwarzschild solution for nonzero coupling constant
$\xi$. As the metric of the solution in the latter family has been constrained
very close to the Schwarzschild metric regardless of the bumblebee
field~\citep{Xu:2022frb}, we have only considered testing solutions in the first
family in this work.   

Using astronomical observations on the shadows and the mass-distance ratios of
two supermassive black holes, we have identified the region for the bumblebee black
holes that are consistent with the observations on the two-dimensional plane of
the coupling constant $\xi$ and the bumblebee charge $Q$ as shown in
Fig.~\ref{fig2}. The parameter space is only moderately constrained. We
attribute this to two features in the solutions of the bumblebee model. First,
we find that the metric of the black-hole solution is exactly the Schwarzschild
metric while the bumblebee charge can be arbitrarily large if $\xi=2\ka$. Thus
for $\xi\sim 2\ka$, the metric is too close to the Schwarzschild metric for the
bumblebee charge to be constrained by the black-hole shadow observations.
Second, for large $|\xi|$, the maximal bumblebee charge that a black hole can
carry is automatically suppressed in the solution, so the metric is expected to
be close to the Schwarzschild metric anyway.  Therefore little parameter space
is excluded for large $|\xi|$ too.

From another point of view, the difficulty in distinguishing a bumblebee black
hole from the Schwarzschild black hole in GR using the observations of
black-hole shadows indicates how capable an alternative to GR the bumblebee
model is. To discriminate between GR and the bumblebee model, 
state-of-the-art techniques of multimessenger and multiwavelength astronomy are desirable, especially observations of GWs and X-rays that encode complementary information about gravity around
black holes in addtion to the radio images from EHT. 
Features of GW polarizations in the bumblebee model
that are very different from those in GR have recently been found in \citet{Liang:2022hxd}, and an estimate of constraints using future observations of extrem mass ratio inspirals (EMRIs) has been carried out in \citet{Liang:2022gdk}. Further work on finding modifications to GWs
emitted in coalescences of stellar-mass black holes would be the next step for putting the
bumblebee model into test against GW observations.
Observations of X-rays from accretion disks of astrophysical black holes also provide exciting opportunities to test black hole spacetime metric. In fact, the spectral analysis techniques of X-rays are now mature enough to give even slightly better constraints on alternative black-hole solutions than those from GWs when elaborate accretion models are set up and suitable sources are carefully chosen as shown in \citet{Tripathi:2020yts} and  \citet{Bambi:2022dtw}. As the current package for analyzing X-ray data from black-hole accretion disks requires analytical spacetime metric solutions \citep{Bambi:2022dtw}, it would be necessary to incorporate algorithms that can use numerical metric solutions into the package to test the numerical bumblebee black holes in the future.
In the meanwhile, with the
next generation EHT (ngEHT) and space-borne sub-millimeter
interferometries~\citep{Blackburn:2019bly, Gurvits:2022wgm}, sharper black hole
images are on the way. A combination of the ngEHT images with X-ray and GW observations
will eventually provide a more complete landscape for bumblebee-like
vector-tensor gravity theories.

\begin{acknowledgments}
We thank Bin Chen and V. Alan Kosteleck\'y for insightful discussions and
comments. This work was supported by the National Natural Science Foundation of
China (11991053, 12147120, 11975027, 11721303), the China Postdoctoral Science
Foundation (2021TQ0018), the National SKA Program of China (2020SKA0120300), the
Max Planck Partner Group Program funded by the Max Planck Society, and the
High-Performance Computing Platform of Peking University. R.X.\ is supported by
the Boya Postdoctoral Fellowship at Peking University.

{\it{Facility}}: EHT
\end{acknowledgments}

\bibliography{refs}{}

\begin{thebibliography}{}
\expandafter\ifx\csname natexlab\endcsname\relax\def\natexlab#1{#1}\fi
\providecommand{\url}[1]{\href{#1}{#1}}
\providecommand{\dodoi}[1]{doi:~\href{http://doi.org/#1}{\nolinkurl{#1}}}
\providecommand{\doeprint}[1]{\href{http://ascl.net/#1}{\nolinkurl{http://ascl.net/#1}}}
\providecommand{\doarXiv}[1]{\href{https://arxiv.org/abs/#1}{\nolinkurl{https://arxiv.org/abs/#1}}}

\bibitem[{Abbott {et~al.}(2019)}]{LIGOScientific:2018mvr}
Abbott, B.~P., {et~al.} 2019, Phys. Rev. X, 9, 031040,
  \dodoi{10.1103/PhysRevX.9.031040}

\bibitem[{Abbott {et~al.}(2021{\natexlab{a}})}]{LIGOScientific:2020ibl}
Abbott, R., {et~al.} 2021{\natexlab{a}}, Phys. Rev. X, 11, 021053,
  \dodoi{10.1103/PhysRevX.11.021053}

\bibitem[{Abbott {et~al.}(2021{\natexlab{b}})}]{LIGOScientific:2021djp}
---. 2021{\natexlab{b}}.
\newblock \doarXiv{2111.03606}

\bibitem[{Abuter {et~al.}(2022)}]{GRAVITY:2021xju}
Abuter, R., {et~al.} 2022, Astron. Astrophys., 657, L12,
  \dodoi{10.1051/0004-6361/202142465}

\bibitem[{Akiyama {et~al.}(2019)}]{EventHorizonTelescope:2019dse}
Akiyama, K., {et~al.} 2019, Astrophys. J. Lett., 875, L1,
  \dodoi{10.3847/2041-8213/ab0ec7}

\bibitem[{Akiyama {et~al.}(2022{\natexlab{a}})}]{EventHorizonTelescope:2022xnr}
---. 2022{\natexlab{a}}, Astrophys. J. Lett., 930, L12,
  \dodoi{10.3847/2041-8213/ac6674}

\bibitem[{Akiyama {et~al.}(2022{\natexlab{b}})}]{EventHorizonTelescope:2022exc}
---. 2022{\natexlab{b}}, Astrophys. J. Lett., 930, L15,
  \dodoi{10.3847/2041-8213/ac6736}

\bibitem[{Akiyama {et~al.}(2022{\natexlab{c}})}]{EventHorizonTelescope:2022urf}
---. 2022{\natexlab{c}}, Astrophys. J. Lett., 930, L16,
  \dodoi{10.3847/2041-8213/ac6672}

\bibitem[{Bailey \& Kosteleck\'y(2006)}]{Bailey:2006fd}
Bailey, Q.~G., \& Kosteleck\'y, V.~A. 2006, Phys. Rev. D, 74, 045001,
  \dodoi{10.1103/PhysRevD.74.045001}

\bibitem[{Bambi(2022)}]{Bambi:2022dtw}
Bambi, C. 2022.
\newblock \doarXiv{2210.05322}

\bibitem[{Blackburn {et~al.}(2019)}]{Blackburn:2019bly}
Blackburn, L., {et~al.} 2019.
\newblock \doarXiv{1909.01411}

\bibitem[{Bluhm(2006)}]{Bluhm:2005uj}
Bluhm, R. 2006, Lect. Notes Phys., 702, 191, \dodoi{10.1007/3-540-34523-X_8}

\bibitem[{Bluhm {et~al.}(2008{\natexlab{a}})Bluhm, Fung, \&
  Kosteleck\'y}]{Bluhm:2007bd}
Bluhm, R., Fung, S.-H., \& Kosteleck\'y, V.~A. 2008{\natexlab{a}}, Phys. Rev.
  D, 77, 065020, \dodoi{10.1103/PhysRevD.77.065020}

\bibitem[{Bluhm {et~al.}(2008{\natexlab{b}})Bluhm, Gagne, Potting, \&
  Vrublevskis}]{Bluhm:2008yt}
Bluhm, R., Gagne, N.~L., Potting, R., \& Vrublevskis, A. 2008{\natexlab{b}},
  Phys. Rev. D, 77, 125007, \dodoi{10.1103/PhysRevD.79.029902}

\bibitem[{Bluhm \& Kosteleck\'y(2005)}]{Bluhm:2004ep}
Bluhm, R., \& Kosteleck\'y, V.~A. 2005, Phys. Rev. D, 71, 065008,
  \dodoi{10.1103/PhysRevD.71.065008}

\bibitem[{Casana {et~al.}(2018)Casana, Cavalcante, Poulis, \&
  Santos}]{Casana:2017jkc}
Casana, R., Cavalcante, A., Poulis, F.~P., \& Santos, E.~B. 2018, Phys. Rev. D,
  97, 104001, \dodoi{10.1103/PhysRevD.97.104001}

\bibitem[{Colladay \& Kosteleck\'y(1997)}]{Colladay:1996iz}
Colladay, D., \& Kosteleck\'y, V.~A. 1997, Phys. Rev. D, 55, 6760,
  \dodoi{10.1103/PhysRevD.55.6760}

\bibitem[{Colladay \& Kosteleck\'y(1998)}]{Colladay:1998fq}
---. 1998, Phys. Rev. D, 58, 116002, \dodoi{10.1103/PhysRevD.58.116002}

\bibitem[{Do {et~al.}(2019)}]{Do:2019txf}
Do, T., {et~al.} 2019, Science, 365, 664, \dodoi{10.1126/science.aav8137}

\bibitem[{Gebhardt {et~al.}(2011)Gebhardt, Adams, Richstone, Lauer, Faber,
  Gultekin, Murphy, \& Tremaine}]{Gebhardt:2011yw}
Gebhardt, K., Adams, J., Richstone, D., {et~al.} 2011, Astrophys. J., 729, 119,
  \dodoi{10.1088/0004-637X/729/2/119}

\bibitem[{Gurvits {et~al.}(2022)}]{Gurvits:2022wgm}
Gurvits, L.~I., {et~al.} 2022, Acta Astronaut., 196, 314,
  \dodoi{10.1016/j.actaastro.2022.04.020}

\bibitem[{Kosteleck\'y(2004)}]{Kostelecky:2003fs}
Kosteleck\'y, V.~A. 2004, Phys. Rev. D, 69, 105009,
  \dodoi{10.1103/PhysRevD.69.105009}

\bibitem[{Kosteleck\'y \& Li(2019)}]{Kostelecky:2018yfa}
Kosteleck\'y, V.~A., \& Li, Z. 2019, Phys. Rev. D, 99, 056016,
  \dodoi{10.1103/PhysRevD.99.056016}

\bibitem[{Kosteleck\'y \& Mewes(2009)}]{Kostelecky:2009zp}
Kosteleck\'y, V.~A., \& Mewes, M. 2009, Phys. Rev. D, 80, 015020,
  \dodoi{10.1103/PhysRevD.80.015020}

\bibitem[{Kosteleck\'y \& Mewes(2012)}]{Kostelecky:2011gq}
---. 2012, Phys. Rev. D, 85, 096005, \dodoi{10.1103/PhysRevD.85.096005}

\bibitem[{Kosteleck\'y \& Mewes(2013)}]{Kostelecky:2013rta}
---. 2013, Phys. Rev. D, 88, 096006, \dodoi{10.1103/PhysRevD.88.096006}

\bibitem[{Kosteleck\'y \& Russell(2011)}]{Kostelecky:2008ts}
Kosteleck\'y, V.~A., \& Russell, N. 2011, Rev. Mod. Phys., 83, 11,
  \dodoi{10.1103/RevModPhys.83.11}

\bibitem[{Liang {et~al.}(2022{\natexlab{a}})Liang, Xu, Lu, \&
  Shao}]{Liang:2022hxd}
Liang, D., Xu, R., Lu, X., \& Shao, L. 2022{\natexlab{a}}, Phys. Rev. D, 106,
  124019, \dodoi{10.1103/PhysRevD.106.124019}

\bibitem[{Liang {et~al.}(2022{\natexlab{b}})Liang, Xu, Mai, \&
  Shao}]{Liang:2022gdk}
Liang, D., Xu, R., Mai, Z.-F., \& Shao, L. 2022{\natexlab{b}}.
\newblock \doarXiv{2212.09346}

\bibitem[{Tasson(2016)}]{Tasson:2016xib}
Tasson, J.~D. 2016, Symmetry, 8, 111, \dodoi{10.3390/sym8110111}

\bibitem[{Tripathi {et~al.}(2021)Tripathi, Zhang, Abdikamalov, Ayzenberg,
  Bambi, Jiang, Liu, \& Zhou}]{Tripathi:2020yts}
Tripathi, A., Zhang, Y., Abdikamalov, A.~B., {et~al.} 2021, Astrophys. J., 913,
  79, \dodoi{10.3847/1538-4357/abf6cd}

\bibitem[{Xu(2023)}]{Xu:2023zjw}
Xu, R. 2023, in {9th Meeting on CPT and Lorentz Symmetry}.
\newblock \doarXiv{2301.12666}

\bibitem[{Xu {et~al.}(2023)Xu, Liang, \& Shao}]{Xu:2022frb}
Xu, R., Liang, D., \& Shao, L. 2023, Phys. Rev. D, 107, 024011,
  \dodoi{10.1103/PhysRevD.107.024011}

\end{thebibliography}
\bibliographystyle{aasjournal}



\end{document}